\documentclass[epj]{svjour}

\usepackage{graphics}

\begin{document}

\title{Finite Temperature Properties of a Supersolid: a RPA Approach}
\subtitle{}
\author{A.J. Stoffel\inst{1,2} \and M. Gul\'{a}csi\inst{1,2}}   
\institute{
Max-Planck-Institute for the Physics of Complex Systems, 
D-01187 Dresden, Germany \and
Nonlinear Physics Centre, Australian National University,
Canberra, ACT 0200, Australia
}

\date{Received: date / Revised version: date}
\abstract{
We study in random-phase approximation the newly discovered supersolid
phase of ${}^4$He and present in detail its finite temperature properties.
${}^4$He is described within a hard-core quantum lattice gas model, with
nearest and next-nearest neighbour interactions taken into account.
We rigorously calculate all pair correlation functions in a cumulant
decoupling scheme. Our results support the importance of the vacancies
in the supersolid phase. We show that in a supersolid the net vacancy
density remains constant as function of temperature, contrary to the
thermal activation theory. We also analyzed in detail the
thermodynamic properties of a supersolid, calculated the jump in the
specific heat which compares well to the recent experiments.
}

\maketitle
\section{Introduction}
One of the biggest accomplishments of theoretical 
condensed matter physics is the ability 
to classify various phases and phase transitions by their mathematical order.
These mathematical orders are usually expressed by order parameters 
reflecting certain limiting behaviours of two particle correlation functions.
This concept makes it possible to describe the properties of 
physically very different systems in a common language and 
establish a link between them. \newline
Without this concept, the counter-intuitive idea of a 
supersolid, i.e. a solid that also exhibits a superflow, 
would not have been conceivable.
In this language supersolidity, firstly  
proposed by Andreev and  Lifshitz\cite{andreev} in 1969 and picked up 
by  Leggett and Chester in 1970\cite{leggett,chester} is the classification
of systems that simultaneously exhibit  diagonal and off-diagonal 
long range order.  Yet the idea of a supersolid  was reluctantly received
by the scientific community, because early experiments
failed to detect any such effect in Helium-4. 
Apart from John Goodkind\cite{goodkind}  who had previously seen suspicious signals
in the ultrasound signals of solid helium, physicists were surprised
when in 2004 Kim and Chan \cite{kim1,kim2} announced the discovery of supersolid
helium. Although equipped with a head start of more than 30 years
the theoretical understanding of the supersolid state lags vastly behind,
not least because supersolidity as we observe it today is nothing like what
the pioneers in early 1970's had anticipated.
\newline
It is now evident that defectons and impurities play a crucial
role in the formation of the supersolid state. However, the data and 
the results of various experiments draw a rather 
complex picture of the supersolid phase. The normal solid
to supersolid transition is not of the usual first or second 
order type transition but bears remarkable resemblance to 
the Kosterlitz Thouless transition well-known in two dimensional 
systems. Furthermore annealing experiments show that 
${}^3$He impurities play a significant role but the data
remain all but conclusive, as the measured superfluid density
varies by order of magnitudes among the different  groups.
Recently a change in the shear modulus of solid helium 
was found and the measured signal almost identically 
mimics the superfluid density measured by Kim and Chan \cite{kim1,kim2}.
Some popular theories give plausible explanations of some aspects of 
the matter. That, such as the vortex liquid theory suggested by Phil Anderson \cite{anderson},
is in good agreement with properties of the phase transition; theories 
based on defection networks are capable of  describing the change in the
shear modulus. However, we feel that a satisfying and comprehensive theory is still 
missing. Part of the problem stems from the complexity of the 
models. Many models are only at sufficient accuracy solvable 
with numerical Monte Carlo methods. The results, doubtlessly 
useful, are seldom intuitive and the lack of analytical
results do not meet our notion of the understanding of a phenomenon.
Other approaches on the other hand are limited in their
analytical significance and rather represent a phenomenological 
ansatz.   
\newline
In this work we attempt to fill a part of this  gap  
 in the theoretical understanding 
of the supersolid phase. We
present a theory of supersolidity in a Quantum-Lattice Gas model (QLG) beyond simple mean-field approaches. 
Following the approach of K.S. Liu and M.E. Fisher\cite{fisher1} we map the QLG model to
the anisotropic Heisenberg model (aHM). 
The method of Green's functions proves to be very successful 
in the description of ferromagnetic and antiferromagnetic phases and 
we use this method to investigate the supersolid phase which 
corresponds to a canted antiferromagnetic phase. 
Applying the Random-phase Approximation (RPA), we broke down 
using cumulant decoupling, the higher order Green's functions
to obtain a closed set of solvable equations.
This method gives a fully quantum mechanical and analytical solution 
of a supersolid phase. We will see that quantum fluctuations have
a significant effect on the net vacancy density of the supersolid 
and we will also see that  the superfluid
state is unstable against zero-energy quasi-particle excitations.
We also derive important thermodynamic properties of this model
and derive formulas for interesting properties such as the 
power law exponents and the jump of the specific heat across
the normal solid to supersolid line.
\newline
The paper is organized as follows: In Section \ref{sec:GenHam}   we
introduce the generic Hamiltonian of a bosonic many 
body system and discretize it to a  
quantum lattice gas model. This model is equivalent 
to the anisotropic Heisenberg model 
in an external field and we will identify the corresponding 
phases in Section \ref{sec:phases}.  In Section \ref{sec:MF} we re-derive the 
classical mean-field solution and discuss briefly their 
significance before we in Section \ref{sec:GF} recapitulate the Green's 
functions for the anisotropic Heisenberg model in the random-phase approximation
at zero temperature. The ground state of the system is 
obtained by solving the corresponding self-consistency 
equation.
In the following two sections we derive basic thermodynamic 
properties as well as ordinary differential equations to calculate
the first and second order phase transition lines.
In Section \ref{sec:expec} we analyse the quasi-particle energy spectrum and 
in the last section we calculate phase diagrams for various
parameter sets and analyse the properties of the supersolid phase and
the corresponding transitions.
\section{Generic Hamiltonian and  Anisotropic Heisenberg Model}\label{sec:GenHam}
If we neglect possible ${}^3$He impurities supersolid Helium-4  is 
a purely bosonic system whose dynamics and structure are 
governed by a generic bosonic Hamiltonian:
\begin{eqnarray} \label{genbosH}
H&=&\int d^3 x \psi^{\dagger}(x)( -\frac{1}{2 m }\nabla^2+
\mu) \psi(x)  \nonumber\\
&&+\frac{1}{2}\int d^3x d^3x'\psi^{\dagger}(x)\psi^{\dagger}(x') V(x-x')
 \psi(x)\psi(x') \nonumber \\
\end{eqnarray}
Here $\psi^{\dagger}(x)$   are the particle creation operators and $\psi(x)$ 
destruction operators and obey the usual bosonic commutation relations.
Hamiltonians such as in Eq. (\ref{genbosH}) are not, due to the vast size 
of the many body Hilbert-space, diagonalisable 
even for simple potentials $V(x)$. The accomplishment of any 
successful theory is to find an approximation that sustains 
the crucial mechanism while reducing the mathematical 
complexity to a traceable level.
Here we follow the method of Matsubara and Matsuda \cite{matsubara}
who successfully introduced the Quantum Lattice Gas (QLG) model
to describe superfluid Helium.
We believe that the quantum lattice gas model is particularly 
useful for supersolids  
as the spatial discretization of  this model
serves as a natural frame for the crystal lattice
of (super)-solid helium and a bipartite 
lattice elegantly 
simplifies the problem of breaking 
translational invariance symmetry for states
that exhibit diagonal long range order. 
\newline
According to Matsubara and Tsuneto \cite{matsuda} the generic 
Hamiltonian Eq. (\ref{genbosH}) in 
the discrete lattice model reads:
\begin{eqnarray}\label{hcbhubb}
H=\mu \sum_i n_i+
\sum_{ij}u_{ij}(a_i^{\dagger}-a_j^{\dagger})
(a_i-a_j)
+\sum_{ij} V_{ij} n_i n_j \nonumber \\
\end{eqnarray}
Here $u_{ij}$ are solely finite  for nearest neighbor and next nearest neighbor hopping
and otherwise zero. The values of $u_{nn}$ and $u_{nnn}$ are  such
that  the kinetic energy is isotropic up to the 4th order.
As atoms do not penetrate each other there 
can exist only one atom at a time on a lattice site 
and consequently
$a^{\dagger}$ and $a$ are the creation and annihilation 
operators of a hard core boson commuting on different lattice sites,
but obey the anti-commutator relations on identical sites.
Equation (\ref{hcbhubb}) is the Hubbard model in 3 dimensions 
for hard core bosons. 
Neither fully bosonic nor fermionic, the lack of
Wicks theorem inhibits the application of 
pertubative field theory though  hard core
bosonic systems are algebraic identical  
to spin systems. A simple transformation,
where the bosonic operators are substituted by 
spin-1/2 operators generating the SU(2) Lie algebra, 
maps the present QLG model onto the anisotropic 
Heisenberg model:
\begin{eqnarray}
H=h^z \sum_i S^z_i+\sum_{ij}J^{\|}_{ij}S^z_iS^z_j+
\sum_{ij}J^{\top}_{ij}(S^x_i S^x_j+S^y_iS^y_j) 
\end{eqnarray}
Here the correspondence between the 
QLG parameters and the spin coupling constants 
is given by:
\begin{eqnarray}
J^{\|}_{ij}=V_{ij}\nonumber \\
J^{\top}_{ij}=-2 u_{ij}
\nonumber\\
h^z=-\mu+\sum_{j}J^{\top}_{ij}
-\sum_{j}J^{\|}_{ij}
\end{eqnarray}
\section{Phases}\label{sec:phases}
\begin{figure}
\centering
\resizebox{5 cm}{!}{
\includegraphics{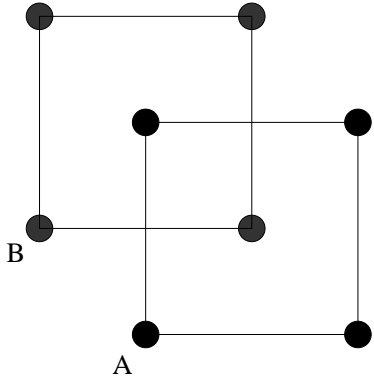} 

}
\caption{\label{fig:lattice} The BCC lattice consists of two interpenetrating 
SC sub-lattices. In the perfect solid phase one sub-lattice (i.e. sub-lattice A) serves
as on-site centers and is occupied while sub-lattice B represents the empty
interstitial and is vacant (For simplicity, we only present the two dimensional
case).}
\end{figure}

The self-consistency equations in the random-phase approximation as we will derive
in a later chapter are very lengthy and therefore it is our
primary goal to keep the present model as simple as
possible while still being able to describe
the crucial physics. For this reason we will define
the anisotropic Heisenberg model on a bipartite
BCC lattice which consists of two interpenetrating
SC sub-lattices as Figure \ref{fig:lattice} shows. In this lattice geometry
the perfectly solid phase is composed of a fully occupied (on-site)
sublattice A while sublattice B refers to the empty interstitial 
and is consequently vacant. As  there is no spatial density variation
 in the liquid phases both sublattices are equally occupied, 
and the mean occupation number simply corresponds to the particle density. 
We are aware that the above choice of SC sublattices does not reflect the
true ${}^4$He crystal structure which is hexagonal closed-packed (HCP). 
Nevertheless we believe that crucial physical properties such as 
phase transition and critical constants do not depend on the specific geometry
as long as no other effects such as frustration are evoked.
\newline
Table \ref{tab:one} charts the various magnetic phases and 
identifies the corresponding phases of the ${}^4$He system. 
According to their spin configurations we identify the four magnetic phases
to be of  
ferromagnetic (FE), canted ferromagnetic (CFE), canted anti-ferromagnetic (CAF) and 
antiferromagntic (AF) orders.
The off-diagonal long range order parameter $m_1$
and the  diagonal long range order $m_2$
are:
\begin{eqnarray}
m_1=\langle S^x_A\rangle+\langle S^x_B\rangle \nonumber \\
m_2=\langle S^z_A\rangle-\langle S^z_B\rangle
\end{eqnarray}
which readily identify the corresponding phases of the Helium system.
\newline\newline
\begin{table}
\label{tab:one}
\caption{ Possible magnetic phases and the corresponding 
phases of the Hubbard model. All Phases are defined by their
long range order. The columns, from left to right, are the spin
configurations, magnetic phases, ODLRO, DLRO and corresponding 
${}^4$He phases, respectively.}
\center
\begin{tabular}{|c|c|c|c|c|}
\hline
&&&&\\
$\uparrow\uparrow$ & FE & No & No & Normal Liquid \\
&&&&\\
$\nearrow\nearrow$ & CFE & Yes & No & Superfluid \\
&&&&\\
$\nearrow\swarrow$ & CAF  & Yes & Yes & Supersolid \\
&&&&\\
$\uparrow\downarrow$& AF & No & Yes & Normal Solid \\
&&&&\\
\hline
\end{tabular}
\end{table}
\section{Mean-Field Solution}\label{sec:MF}
In our previous work \cite{T0paper} we have already re-derived the classical
mean-field solution of the anisotropic Heisenberg model
at zero temperature. As this model provides an easy and intuitive
access to the model we extend the approximation to finite 
temperatures as was done by K.S Liu and M.E. Fisher \cite{fisher1} and briefly
discuss its solution and phase diagram. 
The mean-field Hamiltonian is obtained by substituting the 
spin-$\frac{1}{2}$ operators with their expectation values:
\begin{eqnarray}
 \lefteqn{ H_{MF}=- h^z  (\langle S^z_A \rangle +
    \langle S^z_B \rangle)}\nonumber\\
&&-2 J^{\|}_1 \langle S^z_A\rangle\langle S^z_B\rangle
-J^{\|}_{2}(\langle S^z_A\rangle \langle S^z_A \rangle + 
\langle S^z_B\rangle \langle S^z_B\rangle) \nonumber\\
&&-2 J^{\top}_1 \langle S^x_A\rangle\langle S^x_B\rangle
-J^{\top}_{2}(\langle S^x_A\rangle \langle S^x_A \rangle 
+ \langle S^x_B\rangle \langle S^x_B\rangle ) \nonumber\\
\label{HMF}
\end{eqnarray}
\begin{figure}
\centering
\resizebox{8.5cm}{!}{
\includegraphics*{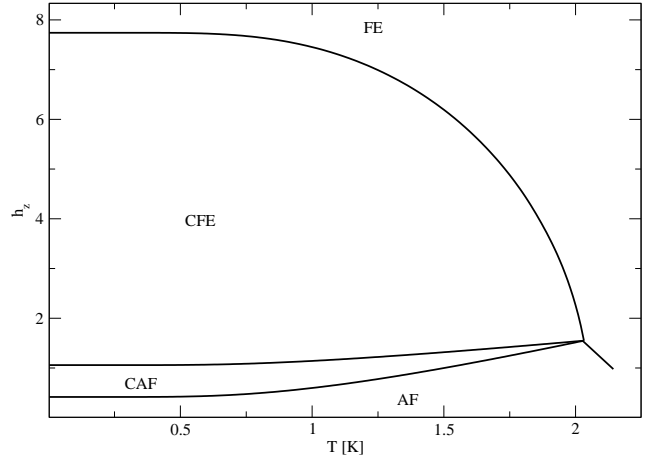}
}
\caption{ The phase
diagram for $J^{\top}_1=1.498$, $J^{\top}_2=0.562$, 
$J^{\|}_1=-3.899$ and $J^{\|}_2=-1.782$ 
as calculated by MF.}
\label{fig:pdmf}
\end{figure}
Here $J^{\|}_1=-q_1 J^{\|}_{i\in A,j\in B} $, $J^{\|}_2=-q_2 J^{\|}_{i\in A,j\in A}$
, $J^{\top}_1=-q_1 J^{\top}_{i\in A,j\in B}$ and $J^{\top}_2=-q_1 J^{\top}_{i\in A,j\in A}$
where $q_1=6$ and $q_2=8$ are the number of nearest and next nearest neighbors.
The mean value of $S_y$ drops out as the randomly broken symmetry
$S_x \leftrightarrow S_y$ (ODLRO) allows for $\langle S_y \rangle=0$. 
The standard method of deriving the corresponding self-consistency 
equations is to minimize the Helmholtz's Free energy $F=H-TS$. 
The entropy S is given by the pseudo spin entropy of the system:
\begin{eqnarray}\label{pspin}
S=-\frac{1}{2}[(\frac{1}{2}+S_A)\ln(\frac{1}{2}+S_A)+(\frac{1}{2}-S_A)\ln(\frac{1}{2}-S_A) \nonumber \\
+(\frac{1}{2}+S_B)\ln(\frac{1}{2}+S_B)+(\frac{1}{2}-S_B)\ln(\frac{1}{2}-S_B)]
\end{eqnarray}
where $S_A=\sqrt{\langle S_{z_A} \rangle^2+\langle S_{x_A}\rangle^2}$ and
 $S_B=\sqrt{\langle S_{z_B}\rangle^2 +\langle S_{x_B}\rangle^2 }$.
In the canted anti-ferromagnetic and the canted ferromagnetic states there are 4 self-consistency equations
in the ferromagnetic and anti-ferromagnetic phases where 
$\langle S_x\rangle=\langle S_y\rangle=0$ they are
reduced by two. These equations are readily obtained by differentiating
 the free energy with respect to $\langle S_z\rangle$ and
 $\langle S_x\rangle$ respectively.
The resulting equations can be rearranged to yield:
\begin{eqnarray}
\sqrt{\langle S_{z_A}\rangle^2+\langle S_{x_A}\rangle^2}=\frac{\tanh(\beta \omega_A)}{2}
\nonumber \\
\sqrt{\langle S_{z_B}\rangle^2+\langle S_{x_B}\rangle^2}=
\frac{\tanh(\beta \omega_B)}{2}\label{mf1}\\
   h^z+2 \langle S_{z_A} \rangle (J_2^{\|}-J_2^{\top})+2 \langle S_{z_B} \rangle J_1^{\|}=
   2 J_1^{\top}\frac{\langle S_{x_B}\rangle }{\langle S_{x_A}\rangle }\langle S_{z_A}\rangle 
   \nonumber \\
   h^z+2 \langle S_{z_B}\rangle  (J_2^{\|}-J_2^{\top})+2 \langle S_{z_A} \rangle J_1^{\|}=
   2 J_1^{\top}\frac{\langle S_{x_A}\rangle }{\langle S_{x_B}\rangle }\langle S_{z_B} \rangle 
   \label{mf2}
\end{eqnarray}
where 
\begin{eqnarray}\label{cmfw}
\omega_A=[(2 J^{\top}_1\langle S^x_B\rangle
          +2 J^{\top}_2 \langle S^x_A \rangle)^2+\nonumber\\
      (2 J^{\|}_1\langle S^z_B\rangle
          +2 J^{\|}_2 \langle S^z_A \rangle+h^z)^2]^{\frac{1}{2}} \nonumber\\
\omega_B=[(2 J^{\top}_1\langle S^x_A\rangle
          +2 J^{\top}_2 \langle S^x_B \rangle)^2+\nonumber\\
      (2 J^{\|}\langle S^z_B\rangle
          +2 J^{\|} \langle S^z_A \rangle+h^z)^2]^{\frac{1}{2}}
\end{eqnarray}
The two equations (Eq. (\ref{mf2}))  are dismissed in the ferromagnetic and the
anti-ferromagnetic phases as the transversal fields $\langle S^x_A \rangle$
and $\langle S^x_B \rangle$ become zero.
At zero temperature $T=0$ the free energy and $\langle H \rangle$ coincide.
This allows us to deduce the phases at absolute zero 
in the limiting cases (limits of $h^z$) from  Equation (\ref{HMF}). 
In the limit of $h^z \rightarrow \infty$ the Hamiltonian reduces to an effective 
one particle model: 
\begin{equation}
H=h^z (\langle S_{z_A}\rangle +\langle S_{z_B}\rangle )
\end{equation}
Consequently the system will be in the energetically favorable
ferromagnetic phase. In the opposite limit, $h^z \rightarrow 0$ and with 
antiferromagnetic coupling $J^{\|}_1<0$ the term of nearest neighbor interaction
\begin{equation}
H=J^{\|}_1 \langle S_{z_A}\rangle\langle S_{z_B}\rangle
\end{equation}
is the only term that significantly lowers the energy. Therefore
the ground-state is the anti-ferromagnetic state. 
At T=0 and a suitable choice of parameters 
the canted ferromagnetic and the canted anti-ferromagnetic phases are realized
in-between these limits as seen in Figure \ref{fig:pdmf}.
However with increasing temperature the regions of 
canted ferromagnetic and canted anti-ferromagnetic phases deplete and
at sufficiently high temperatures only
the anti-ferromagnetic and ferromagnetic phases survive.
As we can see from the phase diagram in Figure \ref{fig:pdmf}  most transitions
are of second order. Only for parameter regimes where the CAF
does not appear the resulting CFE-AF is of first order. 
The boundary lines are defined by ordinary differential equations
and generally need to be calculated numerically. Nonetheless
there exist analytical expressions for the locations of the 
transitions at absolute zero as derived by Matsuda and Tsuneto \cite{matsuda}.
The ferromagnetic to canted ferromagnetic transition point
is determined by Equations (\ref{mf2})   
if we set $\langle S_{z_A}\rangle =\langle S_{z_B}\rangle =\frac{1}{2}$ 
and $\langle S_{x_A}\rangle =\langle S_{x_B}\rangle =0$:
\begin{equation}\label{tpfsf}
h^z_{FE-CFE}=J^{\top}_1+J^{\top}_2-J^{\|}_1-J^{\|}_2
\end{equation}
Equally the 
canted anti-ferromagnetic to anti-ferromagnetic transition
is defined by $\langle S_{z_A}\rangle =-\langle S_{z_B}\rangle =\frac{1}{2}$ 
and $\langle S_{x_A}\rangle =\langle S_{x_B}\rangle =0$:
\begin{equation}\label{tpiaf}
h^z_{CAF-AF}=\sqrt{(-J^{\|}_1+J^{\|}_2-J^{\top}_2)^2-(J^{\top}_1)^2}
\end{equation}
The canted ferromagnetic and the canted anti-ferromagnetic phases 
coexist where the order parameter of the DLRO 
,$m_1=\langle S_{z_A}\rangle-\langle S_{z_B}\rangle $ approaches zero.  
We replace $\langle S_{z_A}\rangle$ and $\langle S_{z_B}\rangle$
 in Equation (\ref{mf2}) with $m_1$ and 
$m_2=\langle S_{z_A}\rangle+\langle S_{z_B}\rangle $
and retain only linear terms of $m_1$. Subtracting and adding both 
equations respectively yields:
\begin{eqnarray}
   h^z+2 m_2(J_2^{\|}-J_2^{\top}+J_1^{\|})=
     2 J_1^{\top}m_2
   \nonumber \\
   2 m_1  (J_2^{\|}-J_2^{\top}- J_1^{\|})=
   -2 J_1^{\top} m1 \frac{ 4m_2^2+1}{4 m_2^2-1}
\end{eqnarray}
We used that $\sqrt{\langle S_{z_A}\rangle^2+\langle S_{x_A}\rangle^2}=\frac{1}{4}$ at T=0.
The solution of these two equations determine
the corresponding transition point which is given by:
\begin{eqnarray}\label{tpisf}
h^z_{CFE-CAF}=\frac{J^{\|}_1+J^{\|}_2-J^{\top}_1-J^{\top}_2}
{-J^{\|}_1+J^{\|}_2+J^{\top}_1-J^{\top}_2}\times\nonumber\\
\sqrt{(-J^{\|}_1+J^{\|}_2-J^{\top}_2)^2-(J^{\top}_1)^2}
\end{eqnarray}
\section{Green's Functions}\label{sec:GF}
Although the classical mean-field theory is quite insightful and gives
an accurate description of the variously ordered phases its fails to take
quantum fluctuations and quasi-particle excitations into account.
Hence, in order to overcome these shortcomings and to derive a fully
quantum mechanical approximation we solve the anisotropic Heisenberg
model in the random-phase approximation which is based on the 
Green's function technique.
At finite temperature the retarded and advanced  Tyablikov \cite{bogo,tyablikov} commutator Green's 
function  defined in real time are:
\begin{eqnarray}\label{GFdef}
G^{\mu\nu}_{ij_{Ret}}(t)=-i\theta(t)\langle |[S^{\mu}_i(t),S^{\nu}_j] | \rangle \nonumber \\
G^{\mu\nu}_{ij_{Adv}}(t)=i\theta(-t)\langle |[S^{\mu}_i(t),S^{\nu}_j] | \rangle
\end{eqnarray}
The average $\langle \rangle$ involves the usual quantum mechanical
as well as thermal averages.  
The most successful technique of solving many body Green's function involves
the method of equation of motion which is given by:
\begin{eqnarray}\label{eq_of_m}
i \partial_t G^{xy}_{ij_{Ret}}(t)= \delta(t)\langle[S^x_i,S^y_j]\rangle
-i\theta(t)\langle[[S^x_i,H],S^y_j]\rangle \nonumber \\
i \partial_t G^{xy}_{ij_{Adv}}(t)= \delta(t)\langle[S^x_i,S^y_j]\rangle
+i\theta(-t)\langle[[S^x_i,H],S^y_j]\rangle \nonumber \\
\end{eqnarray}
The commutator $[S^x_i,H]$ can be eliminated by using the 
Heisenberg equation of motion giving rise to 
higher, third order Green's functions on the RHS. 
In order to obtain a closed set of equations we 
apply the cumulant decoupling procedure which 
splits up the third order differential equation into 
products of single operator expectation values
and two spin Green's functions.
The cumulant decoupling \cite{brown} is based on the assumption that
the last term of the following
equality is negligible:
\begin{eqnarray}
\lefteqn{\langle \hat{A}\hat{B}\hat{C} \rangle =} \nonumber\\
&&\langle \hat{A}\rangle\langle \hat{B}\hat{C} \rangle +
\langle \hat{B}\rangle\langle \hat{A}\hat{C} \rangle \nonumber\\ 
&&+\langle \hat{C}\rangle\langle \hat{A}\hat{B} \rangle 
-2\langle \hat{A}\rangle\langle \hat{B}\rangle\langle\hat{C} \rangle
\nonumber \\
&&+\langle (\hat{A}-\langle \hat{A} \rangle) 
(\hat{B}-\langle \hat{B} \rangle)
(\hat{C}-\langle \hat{C} \rangle) \rangle
\end{eqnarray}
The set of differential equations is not explicitly dependent of the temperature,
i.e. temperature dependence solely comes with thermal averaging of the single 
operator expectation values. Therefore the solution is formally identical to the
zero temperature solution and the detailed derivation of the Green's function
in their full form can be found in Ref. \cite{inprep}. \newline
The averages of the spin operator, appearing in the cumulant
decoupling scheme, have to be determined self-consistently. 
Two self-consistent equations can be derived from correlation
functions corresponding to the Green's functions. 
The self-consistency equations of the canted ferromagnetic (superfluid) and 
canted antiferromagnetic (supersolid) phases 
are structurally different from the ferromagnetic (normal solid) and
antiferromagnetic (normal fluid) 
phases, as the off-diagonal long range order increases the 
number of degrees of freedom by two and hence two additional 
conditions, resulting from the analytical properties of the 
commutator Green's functions, apply.
Thus, the self-consistency equations of the canted phases, 
can be written as a function of the temperature, the 
external magnetic field and the spins in x-direction:
\begin{eqnarray}\label{scs}
F_A^c(\langle S^x_A\rangle ,\langle S^x_B\rangle,h^z,T)=0 \nonumber\\
F_B^c(\langle S^x_A\rangle ,\langle S^x_B\rangle,h^z,T)=0
\end{eqnarray}
similar the self-consistency equations for the ferromagnetic
and anti-ferromagnetic phases:
\begin{eqnarray}\label{scn}
 F_A^{nc}(\langle S^z_A\rangle ,\langle S^z_B\rangle,h^z,T)=0 \nonumber\\
  F_B^{nc}(\langle S^z_A\rangle ,\langle S^z_B\rangle,h^z,T)=0
 \end{eqnarray}
These equations involve a 3 dimensional integral over the 
first Brillouin zone. As explained in the work \cite{T0paper} on the zero temperature 
formalism this integral can be reduced to a two dimensional integral
by introducing a generalized density of state (DOS) and gives the model
a wider applicability.
\section{Thermodynamic Properties}\label{sec:thermo}
The relation between then QLG  and the anisotropic Heisenberg 
model is such that the chemical potential $\mu$ corresponds
to the external field $h^z$, i.e. the grand canonical partition 
function in the QLG corresponds to the canonical partition function
in the anisotropic Heisenberg model where the number of spins is fixed. 
Consequently, thermodynamic potentials of the 
two models are related as:
\begin{equation}
\Theta_{QLG}-\mu N=\Theta_{Heisenberg}
\end{equation}
Here $\Theta$ refers to an arbitrary thermodynamic potential.  \newline
In the same way as the ground state minimizes the internal energy $U=\langle H \rangle$
at absolute zero, the free energy $F=U-T S$ is minimized at finite temperatures.
We wish to stress that the internal energy in the present approximation 
cannot be derived accurately from the expectation value of the Hamiltonian in
the following way: 
\begin{eqnarray}
U=h^z  \sum_i \langle S^z_i \rangle +\sum_{ij}J^{\|}_{ij}\langle S^z_iS^z_j \rangle 
\nonumber \\
+
\sum_{ij}J^{\top}_{ij}(\langle S^x_i S^x_j \rangle +\langle S^y_iS^y_j \rangle )
\end{eqnarray}
The cumulant decoupling, though a good approximation to the anisotropic Heisenberg model, is also an exact 
solution of an unknown effective Hamiltonian $H_{eff}$. Therefore 
thermodynamically consistent results are only obtained if the effective Hamiltonian is 
substituted in the equation above, i.e. $U=\langle H_{eff}\rangle$.
Here, as we do not know the explicit form of this effective model, we have to
integrate the free energy from thermodynamic relations:
\begin{equation}\label{freng}
d  F=\frac{\langle S^z_A\rangle 
+\langle S^z_B\rangle }{2} dh_z+S dT
\end{equation}
This equation allows us to select the ground state in regions
where two or more phases exist self-consistently according 
to the random-phase approximation.
The entropy of the spin system is given by 
\begin{eqnarray}\label{spinentr}
\lefteqn{S=\int d^3k \left(
  \frac{1}{1+\exp(-\beta \omega({\bf{k}}))}\log(\frac{1}{1+\exp(-\beta \omega({\bf{k}}))})\right.
  }\nonumber\\&&
\quad+\left.\frac{1}{1+\exp(\beta \omega({\bf{k}}))}\log(\frac{1}{1+\exp(\beta \omega({\bf{k}}))}) \right)
\end{eqnarray}
\newline
This formula is of purely combinatorial origin and reflects the fact that the hard-core
boson system is equivalent to a fermionic system given by the Jordan-Wigner transformation.
The $\omega(k)$ terms refer to the energies
of the spin-wave excitations. In the solid phases both branches have to be taken into account.
The entropy of the anisotropic Heisenberg model given for a fixed number of spins corresponds to the 
entropy of the QLG model at a constant volume. 
Therefore, in order to obtain the usual configurational entropy of the QLG, we have to
divide by the number of particles per unit cell:
($S_{conf}=\frac{2 S}{n_A+n_B}$). \newline
Here $n_A=1/2-S^z_A$ and $n_B=1/2-S^z_B$ are the particle
 occupation numbers on  lattice sites A and B.
As the nature of the normal solid to supersolid phase transition is not 
yet satisfactorily understood recent experiments \cite{chanspech} have focused on the 
behavior of the specific heat across the transition line in the hope of shedding
light on the matter. 
The specific heat at constant temperature and constant pressure respectively are given by:
\begin{equation}
C_V=T \left( \frac{\partial S_{conf}}{\partial T}\right )_{T,V,N},\quad
C_p=T \left(  \frac{\partial S_{conf}}{\partial T}\right )_{T,P,N}
\end{equation}
Although the external magnetic field $h^z$ in the spin model
is an observable  the corresponding  quantity in the QLG model,
namely the chemical potential is not. Therefore we are also interested in attaining
a formula for the pressure associated with a certain chemical 
potential. 
The relationship is most easily derived from the following Maxwell relation:
\begin{equation}\label{pressa}
\left( \frac{\partial P}{\partial \mu}\right )_{T,V} =\left( \frac{\partial N}
{\partial V}\right )_{T,\mu}=\frac{\#_{lattice\_sites}}{V}(1-\epsilon)
\end{equation}
where $\epsilon:=\langle S^z_A\rangle +\langle S^z_B\rangle$. 
Note that, using this equation in order to obtain the pressure at any 
specific temperature we need a reference point, i.e. 
a chemical potential where the corresponding pressure is known.
This point is given by $\mu \;\rightarrow\; \infty$ which corresponds
to $N\;\rightarrow\;0$ and, hence, $P\; \rightarrow \;0$.
Consequently, in order to obtain the pressure for a specific 
chemical potential $\mu'$ we have to integrate over the interval $[\infty,\mu']$.
\section{Boundary Lines}\label{sec:BL}
The state of the system at any point in $T$ and $h^z$ is given 
by the self-consistency equations and the free energy as
can be derived from Eq. (\ref{freng}). 
Nevertheless the 
resulting computations come at high computational cost 
and therefore it seems most feasible to derive ODEs 
which determine the first and second order transition lines. 
First we will derive the ordinary differential equations which define
the more abundant second order transitions.
The normal fluid (FE) and the normal solid (CFE) phases are
determined by Equations  (\ref{scn}) and the supersolid (CAF) 
and superfluid (CFE) are defined by Equations (\ref{scs}) and condition Equation (\ref{mf2}).
Consequently on the SS-NS and SF-NF transition line, where
both the normal (FE and CFE) and the super (CFE and CAF) 
phases coexists 
following equations hold:
\begin{eqnarray}
F^N_A(\langle S_{z_A}\rangle,\langle S_{z_B}\rangle,h^z,T)\nonumber=0 \\
F^N_B(\langle S_{z_A}\rangle,\langle S_{z_B}\rangle,h^z,T)\nonumber=0 \\
h^z+2 \langle S_{z_A}\rangle (J_2^{\|}-J_2^{\top})+2 \langle S_{z_B}\rangle J_1^{\|}=
   2 J_1^{\top}\frac{\langle S_{x_B}\rangle }{\langle S_{x_A}\rangle }\langle S_{z_A}\rangle
   \nonumber\\
h^z+2\langle S_{z_B}\rangle (J_2^{\|}-J_2^{\top})+2\langle S_{z_A}\rangle J_1^{\|}=
   2 J_1^{\top}\frac{\langle S_{x_A}\rangle }{\langle S_{x_B}\rangle }\langle S_{z_B} \rangle
\end{eqnarray}
On the SS-NS boundary line
the quotient $\langle S_{x_A}\rangle/ \langle S_{z_A}\rangle$ is not known a priori and 
therefore we eliminate it in the equation above yielding:
\begin{eqnarray}
F^N_A(\langle S_{z_A}\rangle,\langle S_{z_B}\rangle,h^z,T)\nonumber=0 \\
F^N_B(\langle S_{z_A}\rangle,\langle S_{z_B}\rangle,h^z,T)\nonumber=0 \\
f(\langle S_{z_A}\rangle,\langle S_{z_B}\rangle,h^z)\nonumber \\
:=(h^z+2 \langle S_{z_A}\rangle (J_2^{\|}(0)-J_2^{\top})+2 \langle S_{z_B}\rangle J_1^{\|})\times
   \nonumber\\
(h^z+2\langle S_{z_B}\rangle (J_2^{\|}-J_2^{\top})+2\langle S_{z_A}\rangle J_1^{\|})-
   (2 J_1^{\top}\langle S_{z_B} \rangle)^2=0
\end{eqnarray}
We introduce a variable s which parametrizes the 
boundary curve. If we for instance choose ds=dT
we get the following set of ordinary differential equations,
defining the NF-SF and the NS-SS transition lines:
\begin{equation}
 \left(\begin{array}{llll} 
\frac{\partial F^N_A}{\partial \langle S_{z_A}\rangle }&
\frac{\partial F^N_A}{\partial \langle S_{z_B}\rangle }&
\frac{\partial F^N_A}{\partial h^z}&
\frac{\partial F^N_A}{\partial T}\\
\frac{\partial F^N_B}{\partial \langle S_{z_A}\rangle }&
\frac{\partial F^N_B}{\partial \langle S_{z_B}\rangle }&
\frac{\partial F^N_B}{\partial h^z}&
\frac{\partial F^N_B}{\partial T}\\
\frac{\partial f}{\partial \langle S_{z_A}\rangle }&
\frac{\partial f}{\partial \langle S_{z_B}\rangle }&
\frac{\partial f}{\partial h^z}&
0\\
0&0&0&1
\end{array}\right)
\cdot
 \left(\begin{array}{l} 
\frac{\partial \langle S_{z_A}\rangle }{\partial s}\\
\frac{\partial \langle S_{z_B}\rangle }{\partial s}\\
\frac{\partial h^z}{\partial s}\\
\frac{\partial T}{\partial s}
\end{array}\right)
=
 \left(\begin{array}{l} 
0\\
0\\
0\\
1
\end{array}\right)
\end{equation}
Upon crossing the SF-SS transition line,
coming from the superfluid phase
the set of possible solutions 
branches off into two phases, the supersolid 
and a non-physical (complex valued) superfluid phase. Therefore
any matrix of ordinary differential equations will render a singularity 
and consequently we have to approach the transition line in 
a limiting process:
\begin{eqnarray}
F^S_A(\langle S_{x_A}\rangle,\langle S_{x_B}\rangle,h^z,T)\nonumber=0 \\
F^S_B(\langle S_{x_A}\rangle,\langle S_{x_B}\rangle,h^z,T)\nonumber=0 \\
\lim_{\epsilon\rightarrow 0}( \langle S_{x_A}\rangle-\langle S_{x_b}\rangle-\epsilon)=0
\end{eqnarray}
The resulting ODE is 
\begin{equation}
 \left(\begin{array}{llll} 
\frac{\partial F^S_A}{\partial \langle S_{x_A}\rangle }&
\frac{\partial F^S_A}{\partial \langle S_{x_B}\rangle }&
\frac{\partial F^S_A}{\partial h^z}&
\frac{\partial F^S_A}{\partial T}\\
\frac{\partial F^S_B}{\partial \langle S_{x_A}\rangle }&
\frac{\partial F^S_B}{\partial \langle S_{x_B}\rangle }&
\frac{\partial F^S_B}{\partial h^z}&
\frac{\partial F^S_B}{\partial T}\\
1&-1&0&0\\
0&0&0&1
\end{array}\right)
\cdot
 \left(\begin{array}{l} 
\frac{\partial \langle S_{x_A}\rangle }{\partial s}\\
\frac{\partial \langle S_{x_B}\rangle }{\partial s}\\
\frac{\partial h^z}{\partial s}\\
\frac{\partial T}{\partial s}
\end{array}\right)
=
 \left(\begin{array}{l} 
0\\
0\\
\epsilon \\
1
\end{array}\right)
\end{equation}
which defines the superfluid to supersolid transition.
As mentioned previously, there are certain parameter regimes where not all four possible phases 
are appearing and consequently a first order transition (mostly superfluid to supersolid)
will occur. In the previous chapter we have seen that such a transition line is 
difficult to locate. However, a tricritical point frequently appears in the 
phase diagram and at this point the first order transition evolves into a second order transition.
This tricritical point can be taken as a initial value for a differential equation 
defining the corresponding first order transition line.\newline
The relevant ODE may be derived from a  Clausius Clapeyron type equation.
On the transition line both phases have equal free energy. Hence:
\begin{eqnarray}
\frac{\Delta S}{\Delta( \langle S_{z_A} \rangle+\langle S_{z_B} \rangle)}=
\frac{\partial T}{\partial h_z}
\end{eqnarray}
where S refers to the spin entropy as derived in the previous section (Eq. (\ref{spinentr}))
and $\Delta$ refers to the difference  of either entropy or spin mean-field
of the superfluid and the normal solid phases. 
This equation together with $ds=dT$ and the total derivative of the  
two self-consistency equations for the normal solid and one equation for the superfluid phase 
form a set of 5 ODEs determining  
$\langle S_x \rangle$ in the superfluid phase
and $  \langle S_{z_A}  \rangle$ and $\langle S_{z_B} \rangle$  in the 
normal solid phase along the boundary line
in the $T-h_z$ plane: 
\begin{equation}\label{fot}
 \left(\begin{array}{lllll} 
\frac{\partial F^S}{\partial \langle S_{x}\rangle }&0&0&
\frac{\partial F^S}{\partial h^z}&
\frac{\partial F^S}{\partial T}\\
0&
\frac{\partial F^N_A}{\partial \langle S_{z_A}\rangle }&
\frac{\partial F^N_A}{\partial \langle S_{z_B}\rangle }&
\frac{\partial F^N_A}{\partial h^z}&
\frac{\partial F^N_A}{\partial T}\\
0&
\frac{\partial F^N_B}{\partial \langle S_{z_A}\rangle }&
\frac{\partial F^N_B}{\partial \langle S_{z_B}\rangle }&
\frac{\partial F^N_B}{\partial h^z}&
\frac{\partial F^N_B}{\partial T}\\
0&0&0&\Delta S&\Delta \langle S_{z_{A+B}} \rangle \\
0&0&0&0&1
\end{array}\right)
\end{equation}
\section{Excitation spectrum} \label{sec:expec}
\begin{figure}
\centering
\resizebox{8.5cm}{!}{
\includegraphics*{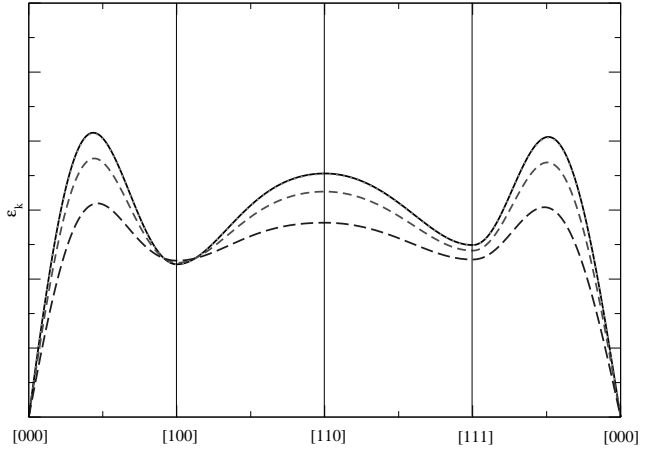} 
}
\caption{\label{fig:disprelsf} 
Excitation spectrum in the superfluid phase 
for for $J^{\top}_1=1.498 K$,
$J^{\top}_2=0.562K$,
$J^{\|}_1=-3.899K$,
$J^{\|}_2=-1.782K$ and
$h_z=3$. Solid  line refers to
$T=0$, dotted to $T=0.5$, dashed
line to $T=1$ and the long dashed line to $T=1.3$.
}
\end{figure}

\begin{figure}
\centering
\resizebox{8.5cm}{!}{
\includegraphics*{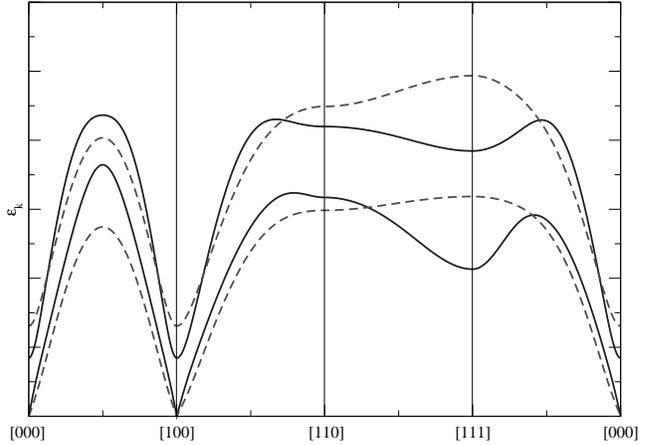} 
}
\caption{\label{fig:disprelss} 
Excitation spectrum in the supersolid phase 
for for $J^{\top}_1=1.498 K$,
$J^{\top}_2=0.562K$,
$J^{\|}_1=-3.899K$,
$J^{\|}_2=-1.782K$ and
$h_z=0.8$. The Solid lines refer to
$T=0$ and the dashed lines to $T=1$.
}
\end{figure}

The superfluid phase features, due to spontaneously 
broken U(1) symmetry, the well know gapless 
Goldstone bosons, i.e. linear phonons. 
The supersolid phase additionally exhibits 
a second, gapped  branch which is due
to the break down of discrete translational symmetry. 
\newline
Figure \ref{fig:disprelss} reveals a zero frequency mode 
in the supersolid phase at [100] of the first Brillouin zone
and consequently the superfluid to supersolid phase transition
is characterized by a collapsing roton minimum at [100].
The dispersion relation in the superfluid (CFE) phase is given by:
\begin{eqnarray}\label{sfdr}
 \omega(k)=2 \{(J^{\top}_1 (\gamma_1(k)-1)+J^{\top}_2 (\gamma_2(k)-1))\times\nonumber\\
  \left[ \langle S_z\rangle^2 (J^{\top}_1
   (\gamma_1(k)-1)+J^{\top}_2 (\gamma_2(k)-1))-\right. \nonumber \\
   \left. \langle S_x\rangle^2 (J^{\top}_1+J^{\top}_2-J^{\|}_1
   \gamma_1(k)-J^{\|}_2 \gamma_2(k)) \right] \}^{1/2}
\end{eqnarray}
From this equation we can see that the energy possibly  goes to 
zero at [100] (corresponds to $\gamma_1(k)=-1$ and $\gamma_2(k)=1$)
when following condition is fulfilled:
\begin{equation}{\label{iscr1}}
J^{\top}_1+J^{\top}_2+J^{\|}_1-J^{\|}_2<0
\end{equation}
Hence we obtain a further condition (supplementary to Eq. (\ref{tpisf}))
for the existence of the superfluid to supersolid transition.
\newline
Equation (\ref{sfdr}) allows for the existence of a second region of the 
reciprocal lattice space where the 
dispersion relation might  go soft. 
For $\gamma_1(k)=0$ and $\gamma_2(k)=-1$ which corresponds to
[111] 
we obtain following condition:
\begin{equation}\label{iscr2}
\frac{-2 J^{\top}_2}{
J^{\top}_1+J^{\top}_2+J^{\|}_2}>0
\end{equation}
It is also interesting to study the behavior of the excitation 
spectrum with increasing temperature. 
In a conventional superfluid the long wave-length behavior 
is given by:
\begin{eqnarray}
\omega(k)=\frac{  n_o (T)V(0)}{m} k
\end{eqnarray}
Here V(0) is the interaction potential at zero momentum and 
m is the particles' mass. The density of the condensate $n_o (T)$
typically decreases with increasing temperature and for that reason we
expect lower energies with increasing temperature. 
In Figure \ref{fig:disprelsf} we can see that the quasi-particle
energies indeed decrease with increasing temperature.
Apart from the region in the vicinity of [100] the energies at 
higher temperatures lie significantly lower than those ones closer
to absolute zero. This is important as it will contribute to the variation
of thermodynamic quantities such as the entropy or the specific heat.
Figure \ref{fig:disprelss} depicts the variation of the 
excitation spectrum with increasing temperature in the supersolid phase.
In this phase the low lying phonon branch mostly depletes with increasing 
temperature, although there exist a region between [110] and [111]
where the zero temperature spectrum is significantly higher. 
Contrary to the phonon branch the gapped mode lifts the energy
around the long wave length limit [000] and around [100].  
In comparison with the superfluid dispersion relation 
the excitation spectrum changes its form/shape
rather than scaling down with increasing temperature as
in the superfluid phase. We observe that the 
supersolid phase exhibits a more complex and 
diverse structure than the superfluid or normal solid
phases alone.
\section{Discussion}
\subsection{On Finite Temperature Properties}
\begin{figure}
\centering
\resizebox{8.5cm}{!}{
\includegraphics*{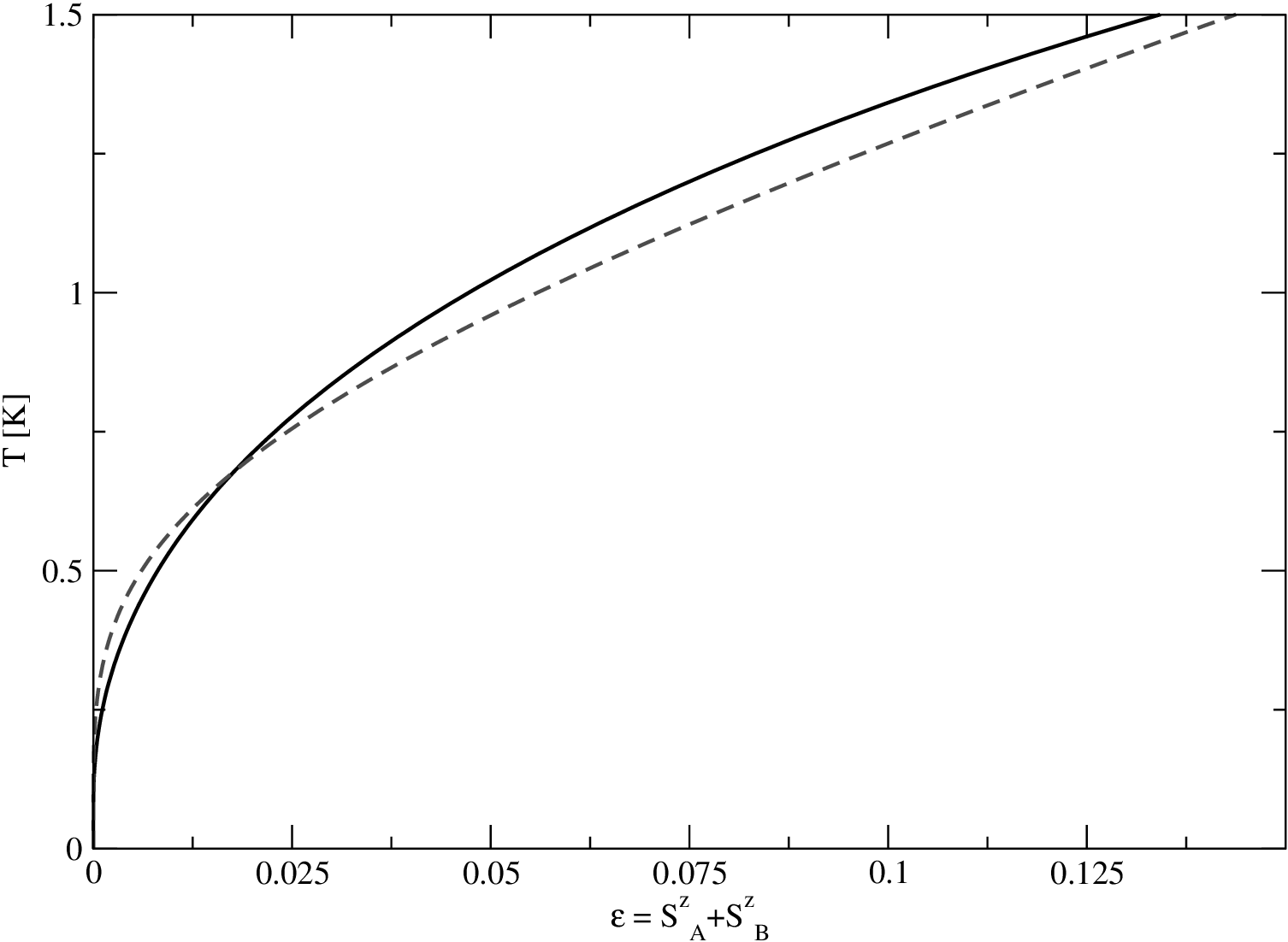} 
}
\caption{\label{fig:epsilon} Curves show the net vacancy density as a 
function of temperature on the supersolid to normal solid transition lines.
The solid line refers to set 1: $J^{\top}_1=1.498 K$,
$J^{\top}_2=0.562 K$,
$J^{\|}_1=-3.899 K$ and
$J^{\|}_2=-1.782 K$  and the dashed line to
set 2: 
$J^{\top}_1=0.5 K$,
$J^{\top}_2=0.5 K$,
$J^{\|}_1=-2.0 K$ and
$J^{\|}_2=-0.5 K$. }
\end{figure}

In this section we will discuss finite temperature properties of the anisotropic 
Heisenberg model and its solution in the random-phase approximation. 
In order to be able to compare the temperature dependence of the model 
in the random-phase approximation with the classical mean-field approximation 
we chose the set of parameters that was extensively scrutinized by Liu and Fisher \cite{fisher1}:
\begin{eqnarray}
J^{\top}_1=1.498 K\nonumber \\
J^{\top}_2=0.562 K\nonumber \\
J^{\|}_1=-3.899 K\nonumber \\
J^{\|}_2=-1.782 K 
\end{eqnarray}
As mentioned in the previous section, Liu and Fisher \cite{fisher1}
have chosen this set of parameters because it provides 
arguably the best fit to the phase diagram of Helium-4.
Since we believe that the validity of the quantum lattice gas model
is too limited to appropriately reproduce the behavior of 
Helium-4 over the whole range of temperature and pressure 
we do not discuss most properties in the pressure-temperature 
space but rather present the major part of the results in the 
more comprehensible magnetic field -temperature coordinates.
Only where the theory can be compared to relevant experimental 
data, such as the heat capacity at constant pressure we 
work in the corresponding representation. 
\begin{figure}
\centering
\resizebox{8.5cm}{!}{
\includegraphics*{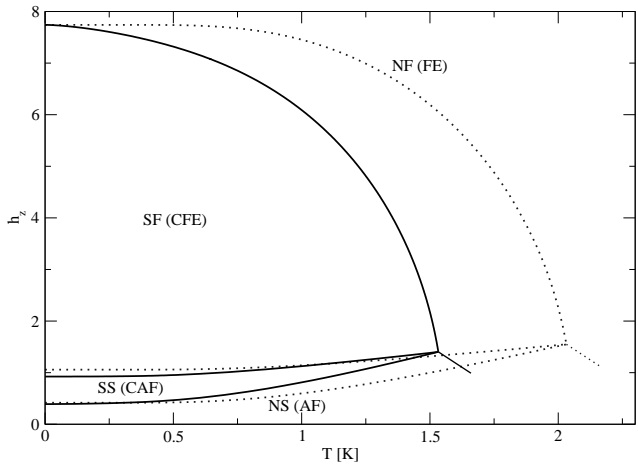} 
}
\caption{\label{fig:pdc1} The phase
diagram for $J^{\top}_1=1.498 K$, $J^{\top}_2=0.562 K$, 
$J^{\|}_1=-3.899 K$ and $J^{\|}_2=-1.782 K$ 
in the RPA (solid lines) and in classical MF (dashed lines).
MF overestimates the temperature by about 30\%.}
\end{figure}
 \begin{figure}
\resizebox{8.5cm}{!}{
\includegraphics*{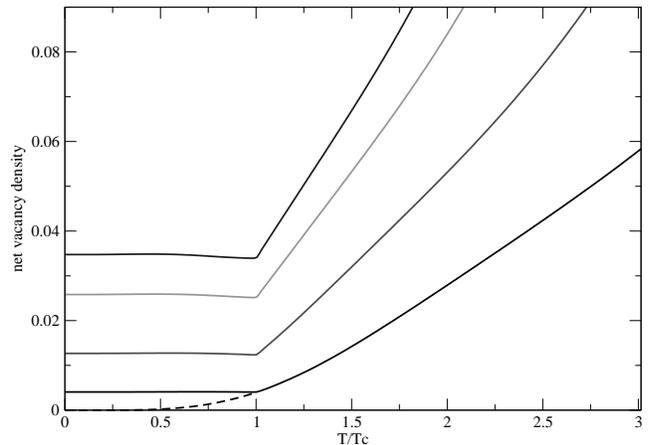} 
}
\caption{\label{fig:nvds} Net vacancy density 
for four different pressures. The dashed line
shows the curve expected if the normal solid would
exist down to zero temperature}
\end{figure}
 \begin{figure}
\centering
\resizebox{8.5cm}{!}{
\includegraphics*{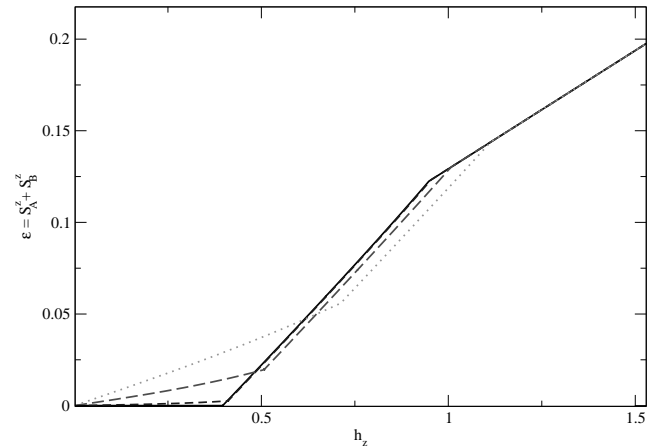} 
}
\caption{\label{fig:isoTsz} Net vacancy density as a function of the
external magnetic field (equates to the chemical potential) for four different temperatures.
The solid line refers to T=0K, the dashed line to T=0.4K, the long dashed line to T=0.7K 
and the dotted line refers to T=1K.}
\end{figure} 
\begin{figure}
\centering
\resizebox{8.5cm}{!}{
\includegraphics*{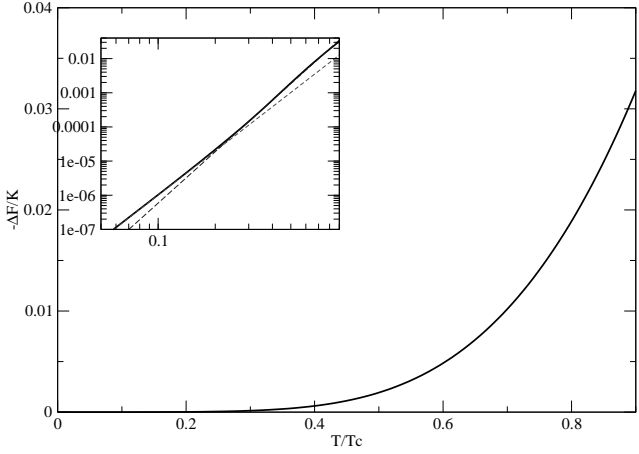} 
}
\caption{\label{fig:freengy} Free energy in the supersolid phase.
The leading contribution comes from a $T^{4.2}$-term. The 
inset shows the free energy in double logarithmic scale. 
The dashed line is the fit to the $T^{4.2}$-term. The 
leading correction comes from a $T^5$-term, indicated 
by the long dashed line.}
\end{figure}
\begin{figure}
\centering
\resizebox{8.5cm}{!}{
\includegraphics*{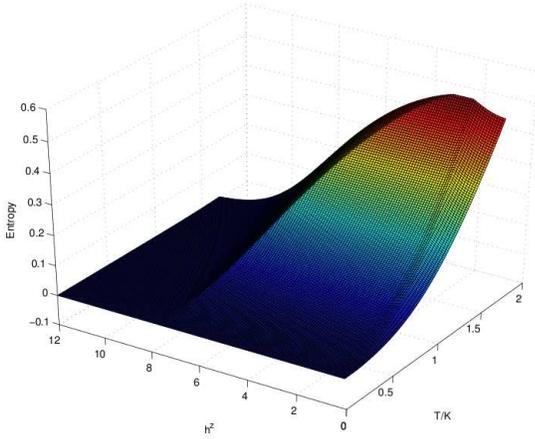} 
}
\caption{\label{fig:entropy}[Color online] 3D plot of the spin entropy. 
The four phases are clearly distinguishable
as the entropy is non-smooth across the transition lines.}
\end{figure}
 \begin{figure}
\centering
\resizebox{8.5cm}{!}{
\includegraphics*{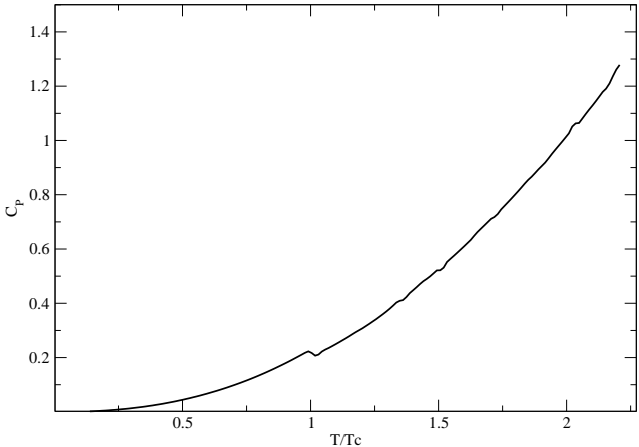} 
}
\caption{\label{fig:specheat} Specific heat at constant pressure.}
\end{figure} 
\newline
The phase diagram of the anisotropic Heisenberg model 
in $T$ and $h^z$ coordinates is given in Figure \ref {fig:pdc1}.
The dashed lines correspond to the phase diagram
of the mean-field approximation. We see that
the diagrams are quantitatively quite similar but in the 
mean-field solution the temperature is somewhat overestimated giving an 
approximately  30\% higher temperature for the 
tetra-critical point.
As mentioned before the critical  magnetic fields $h^z_c$, 
due to quantum fluctuations, are lower in the random-phase approximation.
This effect is most distinctive on the supersolid to superfluid 
transition line as there the deletion of the spin magnitude is 
strongly pronounced. \newline
The net vacancy density $\epsilon$,
density of vacancies minus density of interstitials 
sparked interest as the 
question arose \cite{anderinc} as to whether 
the number of vacancies follow predictions
of thermal activation theory  or are
due to the effects of an incommensurate crystal.
Figure \ref{fig:nvds} shows the net vacancy density
in the supersolid and the normal solid 
phase at constant pressures.  
In isobar curves, curves of constant pressure,
the magnetic field $h^z$ is controlled through Equation (\ref{pressa}). 
We see that the net vacancy density is nearly constant
in the supersolid phase. Only in the 
normal solid phase the net vacancy density increases exponentially
with increasing temperature in total agreement with
classical thermal activation theory. The almost temperature independent 
behavior of the net vacancy density in the supersolid phase is an important 
finding of the quantum lattice gas model and should be observable in 
high resolution experiments if this effect is real.  
In Figure (\ref{fig:isoTsz}) we plotted the net vacancy density as a function of 
$h^z$ (or equivalently $\mu$ the chemical potential) across the 
normal solid, the supersolid and the superfluid phases for 
various temperatures, namely $T=0K$, $T=0.4K$, $T=0.7K$ and $T=1.0K$.
The term net vacancy density does not have a physical meaning in the 
superfluid phase and here the quantity $\epsilon$ rather refers to 
the particle density of the fluid. The particle density in the superfluid phase
is linear in $h^z$ and independent of the temperature $T$, which follows immediately 
from condition Eq. (\ref{mf2}) as $\langle S^x_A\rangle/\langle S^x_B\rangle$ is equal to one 
in the fluid phase. In the supersolid phase the dependence of the net vacancy density on the 
chemical potential is stronger than in the superfluid phase. The chemical potential 
(magnetic field $h^z$) is roughly inversely proportional to the pressure. Meaning that 
the superfluid phase exhibits a higher compressibility than the supersolid phase, which is 
a quite remarkable result. Interestingly the net vacancy density in the supersolid phase 
also increases
linearly  with the chemical potential. This is due to the ratio of the superfluid order parameter
 $\langle S^x_A\rangle/\langle S^x_B\rangle$ varies as the square root of the magnetic field
in the vicinity of the transition:
\begin{eqnarray}
 \langle S^x_A\rangle/\langle S^x_B\rangle \propto 1-\sqrt{h^z-h^z_{CAF-AF}} 
\end{eqnarray}
The exponent $\frac{1}{2}$ is typical for mean-field 
type approximations and appears close to all transition lines. Figure \ref{fig:epsilon}
shows the net vacancy density of the model on the supersolid to normal solid
transition line as a function of temperature, and also reveals the 
mean-field type square root law dependence. \newline 
At zero temperature in the solid phase the net vacancy density is 
equal to zero, hence the crystal is incompressible. In real systems compressibility 
usually occurs as a result of change in the lattice constant a, characterized by the 
Grueneisen parameter. The quantum lattice gas model does not take this effect into account 
since the lattice constant is treated as a constant. However measurements have shown that 
the lattice constant in the (supersolid) helium is almost a constant indicating that the 
net vacancy density is the crucial parameter. 
Also of interest are the free energy 
and the entropy.
Figure (\ref{fig:freengy}) shows the 
free energy of the anisotropic Heisenberg model 
in the supersolid (CAF) phase at constant pressure.
At low T the free energy usually 
follows a power law:
\begin{equation}
F \propto T^{\alpha}
\end{equation}
The coefficient $\alpha$ 
is a universal property of the model 
which is constant over 
the entire regime of a phase. 
The exponent is most easily acquired in the 
log-log plot, shown in the inset of the plot.
In this logarithmic scale $\alpha$ is given by the 
slope of the curve and the leading contribution
is given by, approximately
\begin{eqnarray}
\alpha=4.2
\end{eqnarray}
in the supersolid region. This is close in value to the 
usual $T^4$-term attributed to the linear phonon modes, 
and follows from Equation (\ref{spinentr}) with $\omega(k)$ 
linear in k. Here the $T^4$-term is solely due to the superfluid mode
and in real solids an additional $T^4$-term contribution, accounting 
for the lattice phonon-modes, will appear. The logarithmic plot 
also reveals the leading correction to the free energy given by 
$\alpha=5$.
\newline
The non-configurational entropy of the system 
over the entire range of temperature and magnetic field $h^z$ 
is given in Figure \ref{fig:entropy}. All four phases
are visible and as expected from a thermodynamically
equilibrated system the entropy is monotonically increasing with 
respect to temperature.  \newline
Figure \ref{fig:specheat} depicts the configurational
specific heat at constant pressure in the supersolid and
the normal solid phase. The jump in the specific heat 
at the critical temperature $T_c$, in agreement with the
second order phase transition, appears to be smeared out
due to numerical inaccuracies as the specific heat
is the second derivation of the free energy which 
had to be integrated of the interval $[h^z,\infty]$.
The jump in the specific heat may also be calculated
from following formula which is an analogy to 
the Clausius-Clapeyron equation:
\begin{eqnarray}
 \Delta \frac{\partial^2 F}{\partial {h^z}^2} \left(\frac{d{h^z}^2}{dT^2}\right)_{TL}
+ \Delta \frac{\partial^2 F}{\partial h^z \partial T} \left( \frac{d h^z}{dT}\right)_{TL}
+ \Delta \frac{\partial^2 F}{\partial T^2}
=0\nonumber\\
\end{eqnarray}
\begin{eqnarray}
\lefteqn{\Delta C_h=-T \Delta \frac{\partial^2 F}{\partial T^2}}\nonumber\\&&
=T \Delta \left[ 
\frac{\partial (\langle S_{z_A}\rangle+\langle S_{z_B}\rangle )}{\partial h_z} 
\right]
\left( \frac{\partial h_z}{\partial T} \right)_{TL}^2\nonumber\\&&
+T  \Delta  \left[ \frac{\partial (\langle S_{z_A}\rangle+\langle S_{z_B}\rangle )}{\partial T}  \right]
\left(\frac{\partial h_z}{\partial T}\right)_{TL}
\end{eqnarray}
As we have $C_P=C_h-T \frac{\partial^2 F}{\partial T \partial h_z }
(\frac{\partial h_z}{\partial T})_P$,
we have for the specific heat at constant pressure:
\begin{eqnarray}
\lefteqn{\Delta C_P=T \Delta \left[ 
\frac{\partial (\langle S_{z_A}\rangle+\langle S_{z_B}\rangle )}{\partial h_z} 
\right]
\left( \frac{\partial h_z}{\partial T} \right)_{TL}^2}\nonumber\\&&
+T  \Delta  \left[ \frac{\partial (\langle S_{z_A}\rangle+\langle S_{z_B}\rangle )}{\partial T}  \right]
\left[\left(\frac{\partial h_z}{\partial T}\right)_{TL}-\left(\frac{\partial h_z}{\partial T}\right)_P\right]
\nonumber\\
\end{eqnarray}
For the values corresponding to Figure \ref{fig:specheat} we obtain 
an estimated jump of $0.02$
which is in good agreement with the curve. 
\subsection{First Order Boundary Lines}
\begin{figure}
\centering
\resizebox{8.5cm}{!}{
\includegraphics*{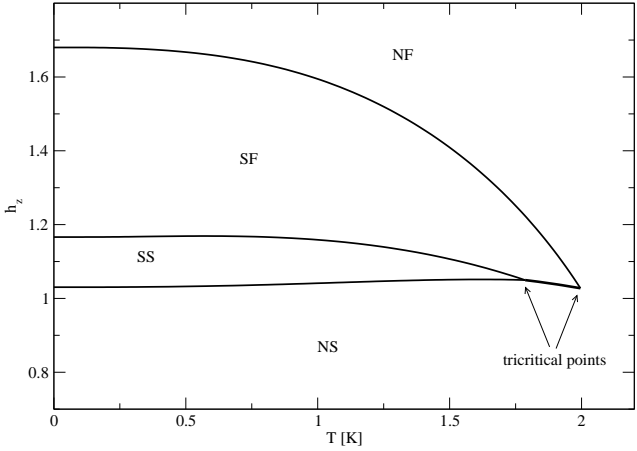} 
}
\caption{\label{fig:c221} The phase
diagram for $J^{\top}_1=0.38 K$, $J^{\top}_2=2.5 K$, 
$J^{\|}_1=-1.2 K$ and $J^{\|}_2=2.4 K$. The supersolid phase 
does not appear above T=1.785K. The two tricritical points are
connected by a first order SF-NS phase transition. }
\end{figure}

In parameter regimes where the supersolid
phase does not appear in certain temperature regions
there consequently appears a first order phase transition between the
superfluid and the normal solid phase. Liu and Fisher \cite{fisher1}
compared the free energies of the competing phases to establish
the transition line. This procedure is not applicable in the random-phase approximation, as 
was outlined in the section on thermodynamic properties. Other than the mean-field approximation
where the Hamiltonian is given by Equation (\ref{HMF}) the effective Hamiltonian 
of the random-phase approximation is not known. 
Therefore we have to integrate the first order transition line from 
a Clausius Clapeyron like equation as derived in the section on Boundary lines.
A set of parameters which exhibits such a first order transition at low temperatures is 
given by:
\begin{eqnarray}
J^{\top}_1=0.5 \nonumber \\
J^{\top}_2=0.5 \nonumber \\
J^{\|}_1=-1.0 \nonumber \\
J^{\|}_2=0.5  
\end{eqnarray}
\begin{figure}
\centering
\resizebox{8.5cm}{!}{
\includegraphics*{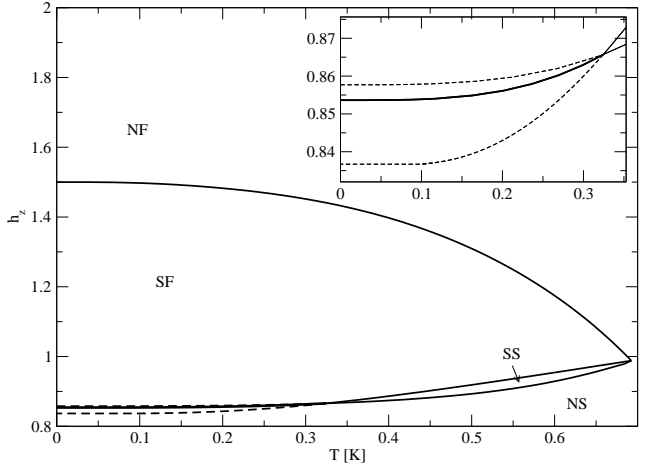} 
}
\caption{\label{fig:c211} The phase
diagram for $J^{\top}_1=0.5 K$, $J^{\top}_2=0.5 K$, 
$J^{\|}_1=-1.0 K$ and $J^{\|}_2=0.5 K$. The supersolid 
phase vanishes below T=0.323K. The resulting SF-NS transition 
is first order.  }
\end{figure}
The corresponding phase diagram is shown in 
Figure \ref{fig:c211}. According to mean-field Eq. (\ref{tpiaf}) and Eq. (\ref{tpisf}) the 
second order superfluid to supersolid  transition as well as
the supersolid to normal solid transition is at absolute zero at $h^z=0.86603$, implying
that the supersolid phase does not exist at zero temperature.
At higher temperatures (above $T>0.323K$) the supersolid phase does exist.
At $T=0.323K$ where the SF-SS and the SS-NS transition lines
intersect there occurs a tricritical point. On this tricritical point the superfluid and normal solid  
phases coexist (as well as the supersolid phase) and the first order phase
transition line can be calculated from the ordinary differential equation as has been derived 
in the section on boundary lines (Eq. (\ref{fot})).
The tricritical point is as such the starting point for the 
integration of the ODE.\newline
There also exist a regime of parameters where
the supersolid phase is suppressed at higher temperatures
as can be seen in Figure \ref{fig:c221} which corresponds to:
\begin{eqnarray}
J^{\top}_1=0.38 \nonumber \\
J^{\top}_2=2.5 \nonumber \\
J^{\|}_1=-1.2 \nonumber \\
J^{\|}_2=2.4 
\end{eqnarray}
The SF-SS and the SS-NS transition lines converge
before the NF-SF line is reached, hence no tetracritical
point such as in Figure \ref{fig:pdc1} is present. The two tricritical points are
connected by a first order superfluid to normal solid transition line.
\subsection{Beyond the Model}
\begin{figure}
\centering
\resizebox{8.5cm}{!}{
\includegraphics*{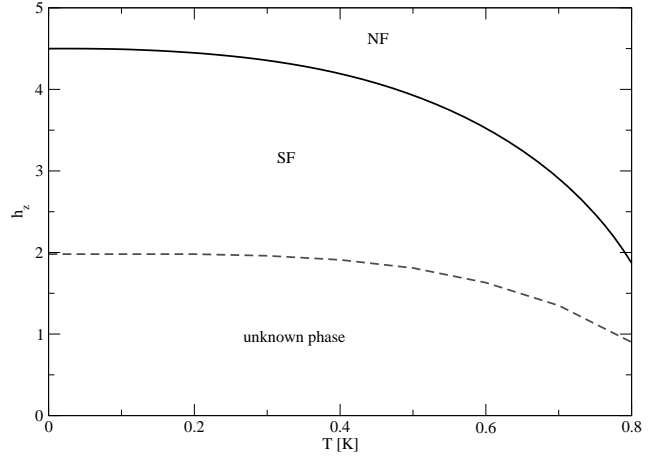} 
}
\caption{\label{fig:pdis2} Phase diagram for $J^{\top}_1=0.5 K$,$J^{\top}_2=0.5 K$,
$J^{\|}_1=-2 K$ and
$J^{\|}_2=-1.5 K$. The superfluid phase becomes unstable due
to a imaginary quasi-particle spectrum (dashed line). RPA does not
yield any stable phase beneath that line.  }
\end{figure}

\begin{figure}
\centering
\resizebox{3.5cm}{!}{
\includegraphics*{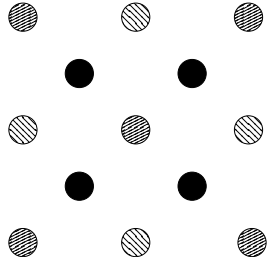} 
}
\caption{\label{fig:sssph}
Two dimensional projection of the lattice structure 
of a (super)-solid phase on three sub-lattices
triggered by a collapse of the roton minimum at [111] of the first
Brillouin zone.
}
\end{figure}

In Section \ref{sec:expec} we have seen that the supersolid phase
appears if and only  if the roton dip at $\gamma_1=-1$ and $\gamma_2=1$, i.e. [001] of the 
first Brillouin zone collapses. Additionally we have seen that the model also allows for a collapsing minimum
at [111] if condition Eq. (\ref{iscr2}) is met. A set of parameters that fulfills this condition is given by:
\begin{eqnarray}
J^{\top}_1=0.5 K\nonumber \\
J^{\top}_2=0.5 K\nonumber \\
J^{\|}_1=-2 K\nonumber \\
J^{\|}_2=-1.5 K  
\end{eqnarray}
Note that the nearest neighbor  and the next nearest neighbor constants in this configuration
$J^{\top}_1$ and $ J^{\top}_2$, corresponding to the kinetic energy, 
 are relatively weak and are both of equal strength, leading to 
a highly anisotropic kinetic energy. 
Figure \ref{fig:pdis2}
shows the corresponding phase diagram according to the random-phase approximation.
The normal fluid to superfluid transition line starts at 
\begin{equation}
h^z=4.5
\end{equation}
at absolute zero and decreases with increasing temperature. 
According to Eq. (\ref{tpiaf}) and Eq. (\ref{tpisf}) the 
critical external fields $h^z$ defining the superfluid to supersolid
and the supersolid to normal solid transitions are, due to negative values
under the square root, imaginary and hence physically not relevant. Consequently in the classical mean-field
approximation 
the superfluid phase extends down to $h^z=0$ and a first order superfluid to 
normal solid transition does not occur as the relatively large negative
$J^{\|}_2$ increases the free energy of a possible solid phase.
\newline
The random-phase approximation however draws a slightly different picture.
Analogous to the classical mean-field solution the random-phase approximation also
yields a phase transition near $h^z=4.5$.
But unlike the classical mean-fields solution, the superfluid phase here does not
survive all the way down to $h^z=0$. 
Due to the particular choice of parameters 
the superfluid phase becomes 
unstable at around $h^z=2$; i.e the quasi-particle spectrum 
turns imaginary at
$\gamma_1(K)=0$ and $\gamma_2(k)=-1$ ([111]). 
Interestingly beyond this line no other 
stable phase exists in the present approach;
there is no set of spin fields $\langle S_{x_A}\rangle$,
 $\langle S_{x_B}\rangle$,
  $\langle S_{z_A}\rangle$ 
and $\langle S_{z_B}\rangle$ that 
solves the self-consistency equations (\ref{scs}) or (\ref{scn}). 
\newline 
To understand why the present approach breaks down and how the 
phase below $h^z=2$ might look like we first investigate the 
physical meaning of the collapsing roton minimum at [100] (equivalently [010] and [001])
leading to the supersolid phase consisting of two SC sub-lattices as already 
analyzed in the previous sections:
The roton dip at [100] corresponds to a density wave given by:
\begin{eqnarray}
\frac{\cos(2\pi x/a)+\cos(2\pi y/a)+\cos(2\pi z/a)}{3}
\end{eqnarray}
This density wave takes on value one on sub-lattice A and minus one on sub-lattice B,
hence it reproduces the periodicity of the supersolid crystal. \newline
In the same way a collapsing roton dip on the main diagonals [111] as given here 
refers to following density wave:
\begin{eqnarray}
\cos(\pi x/a)\cos(\pi y/a)\cos(\pi z/a)
\end{eqnarray}
This density wave yields zero on sub-lattice B and alternatingly
one and minus one on sub-lattice A (neighbors have opposite signs).
Consequently this phase refers to  a (super)-solid phase 
exhibiting three sub-lattices A, B and C where the mean-fields
take on three different values (see Fig. \ref{fig:sssph}). 
It would be quite interesting to study the possibilities and 
properties of such a phase and
we leave as future work the extension of the present approach 
to account for a three sub-lattice phase and
the investigation its properties.
\section{Conclusion}
In this paper we extended the mean-field 
theory of the QLG model by Liu and Fisher \cite{fisher1} at finite temperatures 
and employed a random-phase approximation, where
we derived Green's functions using the method
of equation of motion. We applied the cumulant 
decoupling procedure to split up emergent third order
Green's functions in the EoM.
In comparison to the MF theory employed by Liu and 
Fisher \cite{fisher1} the RPA is a fully quantum mechanically solution 
of the QLG model and therefore takes quantum fluctuations
as well as quasi-particle excitations into account.
The computational results show that these quasi-particle 
excitations are capable to render the superfluid phase unstable 
and thus evoke a phase transition. 
Quantum fluctuations account for additional vacancies and
interstitials even at zero temperature and most interestingly
the net vacancy density is altered in the supersolid phase.\newline
A potential shortcoming of the RPA is the lack of knowledge of 
the effective Hamiltonian and therefore the internal and free energy. 
Usually the free energy is needed to compute first order transition 
lines as the ground state is given by the lowest energy state.
We bypassed this obstacle by deriving a Clausius-Clapeyron like
equation which defines the first order superfluid to normal solid
transition line. The entropy which is an input parameter of this
equation is calculated from the spin wave excitation spectrum.
The jump in the specific heat across the second order 
superfluid to supersolid transition line reveals information about the 
nature of the transition. However the specific heat is the second derivative 
of the free energy and thus the jump is smeared out by the 
numerical calculations. Consequently we derived an equation which 
gives an optional estimation of the jump and is in good agreement with 
the numerical estimate. Most important
our theory predicts a net vacancy density which, in the supersolid phase 
 is significantly different from thermal activation theory. 
In the normal solid phase the net vacancy density
roughly follows the predictions of thermal activation theory, although
quantum mechanical effects give a measurable contribution.
Across the phase transition, however, in the supersolid phase the net vacancy density stays rather
constant as T increases. 
  

\begin{thebibliography}{}    
\bibitem{andreev} A. F. Andreev and I. M. Lifshitz, Zh. Eksp. Teor. Fiz. \textbf{56},(1969) 2057   
[JETP \textbf{29}, (1969) 1107].
\bibitem{chester} G. V. Chester, Phys. Rev. A \textbf{2}, (1970) 256.   
\bibitem{leggett} A. J. Leggett, Phys. Rev. Lett. \textbf{25}, (1970) 1543.  
\bibitem{goodkind}  P. C. Ho, I. P. Bindloss and J. M. Goodkind, J. Low Temp. Physics \textbf{109}, (1997) 409.
\bibitem{kim1} E. Kim, M.H.W. Chan, Nature \textbf{427}, (2004) 225. 
\bibitem{kim2} E. Kim, M.H,W. Chan, Science \textbf{305}, (2004) 1941. 
\bibitem{aoki} Y. Aoki, J. C. Graves, H. Kojima, Phys. Rev. Lett. \textbf{99}, (2007) 015301.   
\bibitem{day} Day, J.R. Beamish, J. Nature \textbf{450}, (2007) 853856. 
\bibitem{anderson} P.W. Anderson, Nature Phys. \textbf{3},(2007) 160.  
\bibitem{T0paper} A. Stoffel and M. Gulacsi, Manuscript submitted to Phys. Rev. B.
\bibitem{inprep} A. Stoffel and M. Gulacsi, Manuscript in preparation.
\bibitem{fisher1} K.-S. Liu, M.E. Fisher, J. Low. Temp. Phys. \textbf{10}, (1973) 655.  
\bibitem{matsuda} H. Matsuda, T. Tsuneto, Prog. Theoret. Phys. Suppl. \textbf{46}, (1970) 411.
\bibitem{matsubara} T. Matsubara and H. Matsuda,Progr. Theoret. Phys. \textbf{16}, (1956) 569;\textbf{17}, (1957).
\bibitem{bogo} N.N. Bogolyubov, S.V. Tyablikov, Doklady Akad. Nauk. S.S.S.R. \textbf{126}, 53 (1959)
 [translation: Soviet Phys. -Doklady \textbf{4},(1959) 604].
\bibitem{tyablikov} S.V. Tyablikov, Ukrain. Mat. Yhur. \textbf{11}, (1959) 287.

\bibitem{brown}P.E. Bloomfield and E.B. Brown, Phys. Rev. B \textbf{22}, (1980) 1353. 
\bibitem{anderinc} P. W. Anderson, W. F. Brinkman and David A. Huse, Science \textbf{310},(2005) 1164.
\bibitem{chanspech} X. Lin, A. Clark and M. Chan, Nature \textbf{449},(2007) 1025.
\end{thebibliography}
\end{document}